 \documentclass[preprint,3p,times,preprintnumber]{elsarticle}

\usepackage{amssymb}
\usepackage{setspace}
\usepackage{amsmath}

\journal{Physics Letters B}

\begin{document} 

\begin{flushright}

IP/BBSR/2016-8

\end{flushright}

\begin{frontmatter}

\title{	Degeneracy between $\theta_{23}$ octant and neutrino non-standard interactions at DUNE}

\address[s1]{Institute of Physics, Sachivalaya Marg, Sainik School Post, Bhubaneswar 751005, India}

\address[s2]{Dipartimento Interateneo di Fisica ``Michelangelo Merlin'', Universit\`a di Bari, Via G.\ Amendola 173, I-70126 Bari, Italy}

\address[s3]{Istituto Nazionale di Fisica Nucleare, Sezione di Bari, Via Orabona 4, 70126 Bari, Italy}

\author[s1]{Sanjib Kumar Agarwalla}
\ead{sanjib@iopb.res.in}

\author[s1]{Sabya Sachi Chatterjee}
\ead{sabya@iopb.res.in}

\author[s2,s3]{Antonio Palazzo}
\ead{palazzo@ba.infn.it}


\begin{abstract}

We expound in detail the degeneracy between the octant of $\theta_{23}$ and 
flavor-changing neutral-current non-standard interactions (NSI's) in neutrino propagation, considering
the Deep Underground Neutrino Experiment (DUNE) as a case study. In the presence of 
such NSI parameters involving the $e-\mu$ ($\varepsilon_{e\mu}$) and $e-\tau$ 
($\varepsilon_{e\tau}$) flavors, the $\nu_\mu \to \nu_e$ and $\bar\nu_\mu \to \bar\nu_e$
appearance probabilities in long-baseline experiments acquire an additional interference term,
which depends on one new dynamical CP-phase $\phi_{e\mu/e\tau}$. 
This term sums up with the well-known interference term related to the standard CP-phase $\delta$ 
creating a source of confusion in the determination of the octant of $\theta_{23}$. We show that 
for values of the NSI coupling (taken one at-a-time) as small as $few\,\%$  
(relative to the Fermi coupling constant $G_{\mathrm F}$), and for unfavorable combinations of the 
two CP-phases $\delta$ and  $\phi_{e\mu/e\tau}$, the discovery potential of the octant of 
$\theta_{23}$ gets completely lost.
 
\end{abstract}


\end{frontmatter}



\section{Introduction}

Although the interactions of neutrinos are well described by the Standard Model (SM) of particle physics,
it is possible that these particles may participate to new non-standard interactions (NSI's),
whose effects are beyond the reach of the existing experiments.
NSI's may appear as a low-energy manifestation of high-energy physics involving new
heavy states (for a review see~\cite{Biggio:2009nt,Ohlsson:2012kf,Miranda:2015dra}) or, 
alternatively, they can be related to new light mediators~\cite{Farzan:2015doa,Farzan:2015hkd}.
As first recognized in~\cite{Wolfenstein:1977ue}, NSI's  can profoundly modify the MSW dynamics~\cite{Wolfenstein:1977ue,Mikheev:1986gs,Mikheev:1986wj} of the neutrino 
flavor conversion in matter. As a consequence, they can be a source of confusion in the
determination of the standard parameters regulating the 3-flavor oscillations if the estimate
of these last ones is extracted from experiments sensitive to MSW effects. Recently, 
in the context of long-baseline (LBL) experiments, the potential confusion between the standard
CP-violation (CPV) related to the 3-flavor CP-phase  $\delta$ and the dynamical CP-phases implied by
neutral-current flavor-changing NSI's has received much attention~\cite{Friedland:2012tq,Rahman:2015vqa,Liao:2016hsa,Forero:2016cmb,Huitu:2016bmb,Bakhti:2016prn,Masud:2016bvp,Soumya:2016enw,deGouvea:2016pom}%
\footnote{Another notable degeneracy occurs  between off-diagonal NSI's and non-zero $\theta_{13}$
in long-baseline~\cite{Huber:2002bi} and solar neutrino experiments~\cite{Palazzo:2009rb,Palazzo:2011vg}. 
Now, this degeneracy has been resolved with the help of  data from reactor experiments (Daya Bay, 
Double Chooz, and RENO), which confirmed that $\theta_{13}$ is non-zero without having any dependency on matter effects.}.

In this paper, we explore in detail, a different kind of degeneracy affecting LBL experiments. It is still 
induced by the new CP-phases related to NSI's, but concerns the octant of the atmospheric mixing angle $\theta_{23}$.
Such a degeneracy has been noted in the  numerical simulations performed in~\cite{Coloma:2015kiu,Blennow:2016etl,deGouvea:2015ndi} and also briefly discussed at the analytical level in~\cite{Liao:2016hsa} (see also~\cite{Friedland:2012tq,Soumya:2016enw}). But, to the best of our knowledge, it has not been addressed in a systematic way
in the literature.
We recall that present global neutrino data~\cite{Capozzi:2016rtj,Gonzalez-Garcia:2015qrr,Forero:2014bxa} indicate
 that $\theta_{23}$ may be non-maximal with two degenerate solutions: 
one $< \pi/4$, dubbed as lower octant (LO), and the other $>\pi/4$, termed as higher octant (HO). Just a few days ago,
at the Neutrino 2016 Conference, the NO$\nu$A collaboration has reinforced the case of two degenerate solutions, excluding maximal mixing at the $2.5 \sigma$ confidence level~\cite{NOvA:2016}. This makes the octant issue even more pressing than before. The identification of the $\theta_{23}$ octant is an important target in neutrino physics, due to the
profound implications for the theory of neutrino masses and mixing (see~\cite{Mohapatra:2006gs,Albright:2006cw,Altarelli:2010gt,King:2014nza,King:2015aea} for reviews).  In the presence of flavor-changing
NSI's involving the $e-\mu$ or $e-\tau$ sectors, the $\nu_\mu \to \nu_e$ transition probability probed at LBL facilities
acquires a new interference term that depends on one new dynamical CP-phase $\phi$. This term sums up with the well-known interference term related to the standard CP-phase $\delta$ creating a potential source of confusion 
in the reconstruction of the $\theta_{23}$ octant. 
Taking the Deep Underground Neutrino Experiment (DUNE)~\cite{Acciarri:2016crz,Acciarri:2015uup,Strait:2016mof,Acciarri:2016ooe,Adams:2013qkq} as a case study,%
\footnote{Recent work on the impact of NSI's at DUNE can be found in~\cite{deGouvea:2015ndi,Coloma:2015kiu,Blennow:2016etl,Masud:2016bvp,deGouvea:2016pom,Bakhti:2016gic}.}
we show that for values of the NSI coupling as small as $few\,\%$  (relative to the Fermi constant $G_{\mathrm F}$), for unfavorable combinations of the two CP-phases $\delta$ and $\phi$, the discovery potential of the octant of $\theta_{23}$ gets completely lost.

\section{Theoretical framework}

A neutral-current NSI can be described by a four-fermion dimension-six operator~\cite{Wolfenstein:1977ue}
\begin{equation}
\mathcal{L}_{\mathrm{NC-NSI}} \;=\;
-2\sqrt{2}G_F 
\varepsilon_{\alpha\beta}^{fC}
\bigl(\overline{\nu_\alpha}\gamma^\mu P_L \nu_\beta\bigr)
\bigl(\overline{f}\gamma_\mu P_C f\bigr)
\;,
\label{H_NC-NSI}
\end{equation}
where subscripts $\alpha, \beta = e,\mu,\tau$ indicate the 
neutrino flavor, superscript $f = e,u,d$ labels the matter 
fermions, superscript $C=L, R$ denotes the chirality of the 
$ff$ current, and $\varepsilon_{\alpha\beta}^{fC}$ are 
dimensionless quantities which parametrize the strengths 
of the NSI's. The hermiticity of the interaction demands
\begin{equation}
\varepsilon_{\beta\alpha}^{fC} \;=\; (\varepsilon_{\alpha\beta}^{fC})^*
\;.
\end{equation}
For neutrino propagation through matter, the relevant combinations are
\begin{equation}
\varepsilon_{\alpha\beta}
\;\equiv\; 
\sum_{f=e,u,d}
\varepsilon_{\alpha\beta}^{f}
\dfrac{N_f}{N_e}
\;\equiv\;
\sum_{f=e,u,d}
\left(
\varepsilon_{\alpha\beta}^{fL}+
\varepsilon_{\alpha\beta}^{fR}
\right)\dfrac{N_f}{N_e}
\;,
\label{epsilondef}
\end{equation}
where $N_f$ denotes the number density of fermion $f$.
For the Earth, we can assume neutral and isoscalar matter, implying  $N_n \simeq N_p = N_e$, 
in which case $N_u \simeq N_d \simeq 3N_e$.
Therefore,
\begin{equation}
\varepsilon_{\alpha\beta}\, \simeq\,
\varepsilon_{\alpha\beta}^{e}
+3\,\varepsilon_{\alpha\beta}^{u}
+3\,\varepsilon_{\alpha\beta}^{d}
\;.
\end{equation}
The NSI's  modify the effective Hamiltonian for neutrino propagation 
in matter, which in the flavor basis reads
\begin{equation}
H \;=\; 
U
\begin{bmatrix} 
0 & 0 & 0 \\ 
0 & k_{21}  & 0 \\ 
0 & 0 & k_{31} 
\end{bmatrix}
U^\dagger
+
V_{\mathrm{CC}}
\begin{bmatrix}
1 + \varepsilon_{ee}  & \varepsilon_{e\mu}      & \varepsilon_{e\tau}   \\
\varepsilon_{e\mu}^*  & \varepsilon_{\mu\mu}    & \varepsilon_{\mu\tau} \\
\varepsilon_{e\tau}^* & \varepsilon_{\mu\tau}^* & \varepsilon_{\tau\tau}
\end{bmatrix}\,,
\end{equation}
where $U$ is the Pontecorvo-Maki-Nakagawa-Sakata (PMNS) matrix, which, in the standard parameterization,
depends on three mixing angles ($\theta_{12}, \theta_{13}, \theta_{23}$) and  one CP-phase ($\delta$).
We have also introduced the solar and atmospheric wavenumbers $k_{21} \equiv \Delta m^2_{21}/2E$ 
and  $k_{31} \equiv \Delta m^2_{31}/2E$ and the charged-current matter potential 
\begin{equation}
V_{\mathrm{CC}} 
\;=\; \sqrt{2}G_F N_e 
\;\simeq\; 7.6\, Y_e \times 10^{-14}
\bigg[\dfrac{\rho}{\mathrm{g/cm^3}}\bigg]\,\mathrm{eV}\,,
\label{matter-V}
\end{equation}
where $Y_e = N_e/(N_p+N_n) \simeq 0.5$ is the relative electron number density in the Earth crust.
It is useful to introduce the dimensionless quantity $v = V_{\mathrm{CC}}/k_{31}$, whose absolute value
is given by
\begin{equation}
|v| 
\;=\; \bigg|\frac{V_{\mathrm{CC}}}{k_{31}}\bigg| 
\;\simeq\; 0.22 \bigg[\frac{E}{2.5\, \mathrm{GeV}}\bigg]\;,
\label{matter-v}
\end{equation}
since it will appear in the analytical expressions of the $\nu_\mu \to \nu_e$ transition probability.
In Eq. (\ref{matter-v}), we have taken the energy of the DUNE first oscillation maximum $E = 2.5\, {\mathrm{GeV}}$ as
a benchmark.

In the present work, we limit our investigation to flavor non-diagonal NSI's, that is, we only allow the 
$\varepsilon_{\alpha\beta}$'s with $\alpha\ne\beta$ to be non-zero.
More specifically, we will focus our attention on the couplings $\varepsilon_{e\mu}$
and $\varepsilon_{e\tau}$, which, as will we discuss in detail, introduce an observable dependency 
from their associated CP-phase in the appearance $\nu_\mu \to \nu_e$ probability
probed at the LBL facilities. For completeness, we will comment about the (different) 
role of the third coupling $\epsilon_{\mu\tau}$, which mostly affects
the $\nu_\mu \to \nu_\mu$ disappearance channel  and has not a critical impact
in the $\theta_{23}$ octant reconstruction. We recall that the current upper bounds (at 90\% C.L.) on the two NSI's under consideration are: $|\varepsilon_{e\mu}| \lesssim 0.33$, as reported in the review~\cite{Biggio:2009nt}, and $|\varepsilon_{e\tau}| \lesssim 0.45$ as derived from the most recent Super-Kamiokande atmospheric data analysis~\cite{Mitsuka:2011ty} under the assumption $\epsilon_{ee} = 0$ (see also~\cite{Gonzalez-Garcia:2013usa}).
 As we will show in detail, the strengths of $|\varepsilon_{e\mu}|$ and $|\varepsilon_{e\tau}|$ that can give rise
to a degeneracy problem with the octant of $\theta_{23}$ are one order of magnitude smaller than these upper
bounds.

\section{Analytical Expressions}

Let us consider the transition probability relevant for the LBL experiment DUNE.
Using the expansions available in the literature~\cite{Kikuchi:2008vq} one can see 
that in the presence of a NSI, the transition probability can be written approximately as 
the sum of three terms
\begin{eqnarray}
\label{eq:Pme_4nu_3_terms}
P_{\mu e}  \simeq  P_{\rm{0}} + P_{\rm {1}}+   P_{\rm {2}}\,,
\end{eqnarray}
where the first two terms return the standard 3-flavor probability and the
third one is ascribed to the presence of NSI.
Noting that the small mixing angle $\sin \theta_{13}$,
the matter parameter $v$ and the modulus $|\varepsilon|$ of the NSI can be considered approximately
of the same order of magnitude $\mathcal{O}(\epsilon)$, while $\alpha \equiv \Delta m^2_{21}/ \Delta m^2_{31} = \pm 0.03$ is
$\mathcal{O}(\epsilon^2)$, one can expand the probability keeping only third order terms. Using a notation
similar to that adopted in ~\cite{Liao:2016hsa}, we obtain%
\footnote{Interestingly, a similar decomposition of the transition probability is valid in the presence of 
a light sterile neutrino~\cite{Klop:2014ima}. In that case, however, the origin of the new interference
term $P_2$ is kinematical, and it is operative also in vacuum. In fact, the new term is related to 
the interference of the atmospheric oscillations with those induced by the new large mass-squared 
splitting implied by the sterile state.}
\begin{eqnarray}
\label{eq:P0}
 & P_{\rm {0}} &\!\! \simeq\,  4 s_{13}^2 s^2_{23}  f^2\,,\\
\label{eq:P1}
 & P_{\rm {1}} &\!\!  \simeq\,   8 s_{13} s_{12} c_{12} s_{23} c_{23} \alpha f g \cos({\Delta + \delta})\,,\\
 \label{eq:P2}
 & P_{\rm {2}} &\!\!  \simeq\,  8 s_{13} s_{23} v |\varepsilon|   
 [a f^2 \cos(\delta + \phi) + b f g\cos(\Delta + \delta + \phi)]\,,
\end{eqnarray}
where $\Delta \equiv  \Delta m^2_{31}L/4E$ is the atmospheric oscillating frequency related
to the baseline $L$.  For compactness, we have used the notation
($s_{ij} \equiv \sin \theta_{ij} $, $c_{ij} \equiv \cos \theta_{ij}$), and following~\cite{Barger:2001yr},
we have introduced the quantities
\begin{eqnarray}
\label{eq:S}
f \equiv \frac{\sin [(1-v) \Delta]}{1-v}\,, \qquad  g \equiv \frac{\sin v\Delta}{v}\,.
\end{eqnarray}
We observe that $P_{\rm {0}}$ is positive definite being independent of the CP-phases, and gives the leading contribution to the probability. In $P_{\rm {1}}$ one recognizes the standard 3-flavor interference term between the solar and the atmospheric frequencies. The third term $P_{\rm {2}}$ brings the dependency on the (complex) NSI coupling and of course is non-zero only in the presence of matter (i.e. if $v \ne 0$). In Eq.~(\ref{eq:P2}) we have assumed for
the NSI coupling the general complex form
\begin{eqnarray}
\varepsilon = |\varepsilon |  e^{i\phi}\,.
\end{eqnarray}
The expression of $P_2$ is slightly different for  $\varepsilon_{e\mu}$ and  $\varepsilon_{e\tau}$ and,
in Eq. (\ref{eq:P2}), one has to put
\begin{eqnarray}
 \label{eq:P2_NSI}
 a = s^2_{23}, \quad b = c^2_{23} \qquad &{\mathrm {if}}& \qquad \varepsilon = |\varepsilon_{e\mu}|e^{i{\phi_{e\mu}}}\,,\\
 a =  s_{23}c_{23}, \quad b = -s_{23} c_{23} \qquad &{\mathrm {if}}& \qquad \varepsilon = |\varepsilon_{e\tau}|e^{i{\phi_{e\tau}}}\,.
\end{eqnarray}
In the expressions given above for $P_0$, $P_1$ and $P_2$, one should bear in mind that
the sign of $\Delta$, $\alpha$ and $v$ is positive (negative) for NH (IH).  In addition, we stress that
the expressions above are valid for neutrinos, and that the corresponding ones for antineutrinos 
are obtained by inverting the sign of all the CP-phases, and of the dimensionless quantity $v$. 

Now let us come to the $\theta_{23}$ octant issue. As a first step it is useful to quantify 
the size of the perturbation from maximal mixing allowed by current data. We can express
the atmospheric mixing angle as 
\begin{eqnarray}
 \label{eq:theta_23}
\theta_{23}  = \frac{\pi}{4} \pm \eta\,,
\end{eqnarray}
where $\eta$ is a positive-definite angle. The positive (negative) sign
corresponds to HO (LO). The current 3-flavor global analyses~\cite{Capozzi:2016rtj,Gonzalez-Garcia:2015qrr,Forero:2014bxa} indicate that $\theta_{23}$ cannot deviate from $45^0$ by more than $\sim 6^0$, i.e.
 $s^2_{23}$ must be in the range $\sim [0.4,0.6]$. Therefore, one has $\eta \lesssim 0.1$, 
and we can use the expansion 
\begin{eqnarray}
 \label{eq:s2_theta_23}
s^2_{23} =  \frac{1}{2} (1 \pm \sin 2\eta\,)   \simeq \frac{1}{2} \pm \eta\,.
\end{eqnarray}
An experiment is sensitive to the octant if, in spite of the freedom provided by the unknown 
CP-phases, there is still a non-zero difference among the transition probability in the two octants, i.e.
\begin{eqnarray}
\label{eq:DPme}
\Delta P \equiv P^{\mathrm {HO}}_{\mu e} (\theta_{23}^{\mathrm {HO}}, \delta^{\mathrm {HO}}, \phi^{\mathrm {HO}}) -
                           P^{\mathrm {LO}}_{\mu e} (\theta_{23}^{\mathrm {LO}}, \delta^{\mathrm {LO}}, \phi^{\mathrm {LO}})\ne 0\,.
\end{eqnarray}
In Eq.~(\ref{eq:DPme}) one of the two octants should be thought as the true octant
(whose value is used to simulate the data) and the other one as the test one (whose value is 
used to simulate the theoretical model).
For example, if for definiteness we fix the HO as the true octant, then
for a given combination of $(\delta^{\mathrm {HO}}_{\mathrm{true}}, \phi^{\mathrm {HO}}_{\mathrm{true}}$)
there is sensitivity to the octant if  there exist  some values of the test phases
$(\delta^{\mathrm {LO}}_{\mathrm{test}}, \phi^{\mathrm {LO}}_{\mathrm{test}})$
such that $\Delta P \ne 0$ at a detectable level%
\footnote{In the numerical analysis the values of the test parameters are determined by minimization
of the $\Delta \chi^2$ [see Eq.~(\ref{eq:chi2}) in Section 4].}.

 According to Eq.~(\ref{eq:Pme_4nu_3_terms}), 
we can split  $\Delta P$ in the sum of three terms 
\begin{eqnarray}
\label{eq:DPme_4nu_3_terms}
\Delta P =   \Delta P_{\rm{0}} + \Delta P_{\rm {1}} +   \Delta P_{\rm {2}}\,.
\end{eqnarray}
The first term is positive-definite, does not depend on the CP-phases and, at the first order in $\eta$,
it is given by 
\begin{eqnarray}
\label{eq:DP_0}
\Delta P_{\rm {0}} \simeq 8 \eta s_{13}^2 f^2\,.
\end{eqnarray}
The second and third terms depend on the CP-phases and can have both positive
or negative values. Their expressions are given by
\begin{eqnarray}
\label{eq:DP_1}
\Delta P_{\rm {1}}& =& A \big[ \cos(\Delta + \delta^{\mathrm {HO}}) - \cos(\Delta + \delta^{\mathrm {LO}})\big] \,,\\
\label{eq:DP_2}
\Delta P_{\rm {2}} &= &
B \big[\cos(\delta^{\mathrm{HO}} + \phi^{\mathrm{HO}}) - \cos(\delta^{\mathrm{LO}} + \phi^{\mathrm{LO}})\big] \pm
C \big[\cos(\Delta + \delta^{\mathrm {HO}} + \phi^{\mathrm{HO}}) - \cos(\Delta + \delta^{\mathrm {LO}} + \phi^{\mathrm{LO}})\big]\,,
\end{eqnarray}
where for compactness, we have introduced the amplitudes%
\footnote{In the expressions of $A$, $B$ and $C$ we are neglecting terms proportional to powers 
of $\eta$, which would give rise to negligible corrections.}
\begin{eqnarray}
\label{eq:A}
A & =&  4 s_{13} s_{12} c_{12} \alpha fg\,,\\
\label{eq:B}
B &= &  2 \sqrt{2} v |\varepsilon| s_{13} f^2 \,,\\
\label{eq:C}
C &= &  2 \sqrt{2} v |\varepsilon| s_{13} f g\,.
\end{eqnarray}
The positive (negative) sign in front of the coefficient $C$ in Eq. (\ref{eq:DP_2}) corresponds
to $\varepsilon_{e\mu}$ ($\varepsilon_{e\tau}$).
In order to get a feeling of the size of the three terms of $\Delta P$
we provide a ballpark estimate adopting as a benchmark case ($\nu$, NH), 
and fixing the energy at the value $E = 2.5$\,GeV corresponding to the first oscillation maximum ($\Delta = \pi/2$), 
in which case $|f| = |\sin \Delta| = 1$ and $|g| = |\Delta| = \pi/2$. For the 3-flavor parameters, we have
used the values provided at the beginning of the next section. For the first term we find
\begin{eqnarray}
\label{eq:DP0_num}
\Delta P_{\rm {0}} \simeq   1.26 \times 10^{-2} \bigg[\frac{\eta}{0.05}\bigg]\,,
\end{eqnarray}
where we have left manifest the linear dependency on the deviation $\eta$ from maximal $\theta_{23}$. 
 The amplitude of the standard interference term $\Delta P_1$ is
\begin{eqnarray}
\label{eq:A}
|A| \simeq   1.52 \times 10^{-2}\,, 
\end{eqnarray}
while, for the two coefficients entering the NSI-induced term $\Delta P_2$, one finds
\begin{eqnarray}
\label{eq:BC}
|B| + |C| \simeq 1.51 \times 10^{-2} \bigg[\frac{|\varepsilon|}{0.05}\bigg]\,,
\label{eq:A_sin}
\end{eqnarray}
where we have left evident the linear dependency on the NSI strength $|\varepsilon|$. 
From this last relation we see that for values of the NSI coupling $|\varepsilon | \sim 0.05$, 
the difference induced by the new interference term ($\Delta P_2$) has approximately 
the same amplitude of that arising form the standard interference term ($\Delta P_1$). 
Also, it is essential to notice that the third term $\Delta P_2$ in Eq. (\ref{eq:DP_2}) depends not only 
on the standard CP-phase $\delta$ but also on the new dynamical CP-phase $\phi$ related to the NSI. 
Since the two CP-phases $\delta$ and $\phi$ are independent quantities, in the SM+NSI scheme there is much
more freedom with respect to the SM case, where only one phase ($\delta$) is present. Therefore, for sufficiently
large values of the NSI coupling, it is reasonable to expect a degradation of the reconstruction of
the $\theta_{23}$ octant, which will depend on the amplitude of the deviation $\eta$.

\begin{figure}[t]
\vspace{-0.5cm}
\hspace{1cm}
\includegraphics[width=14.cm]{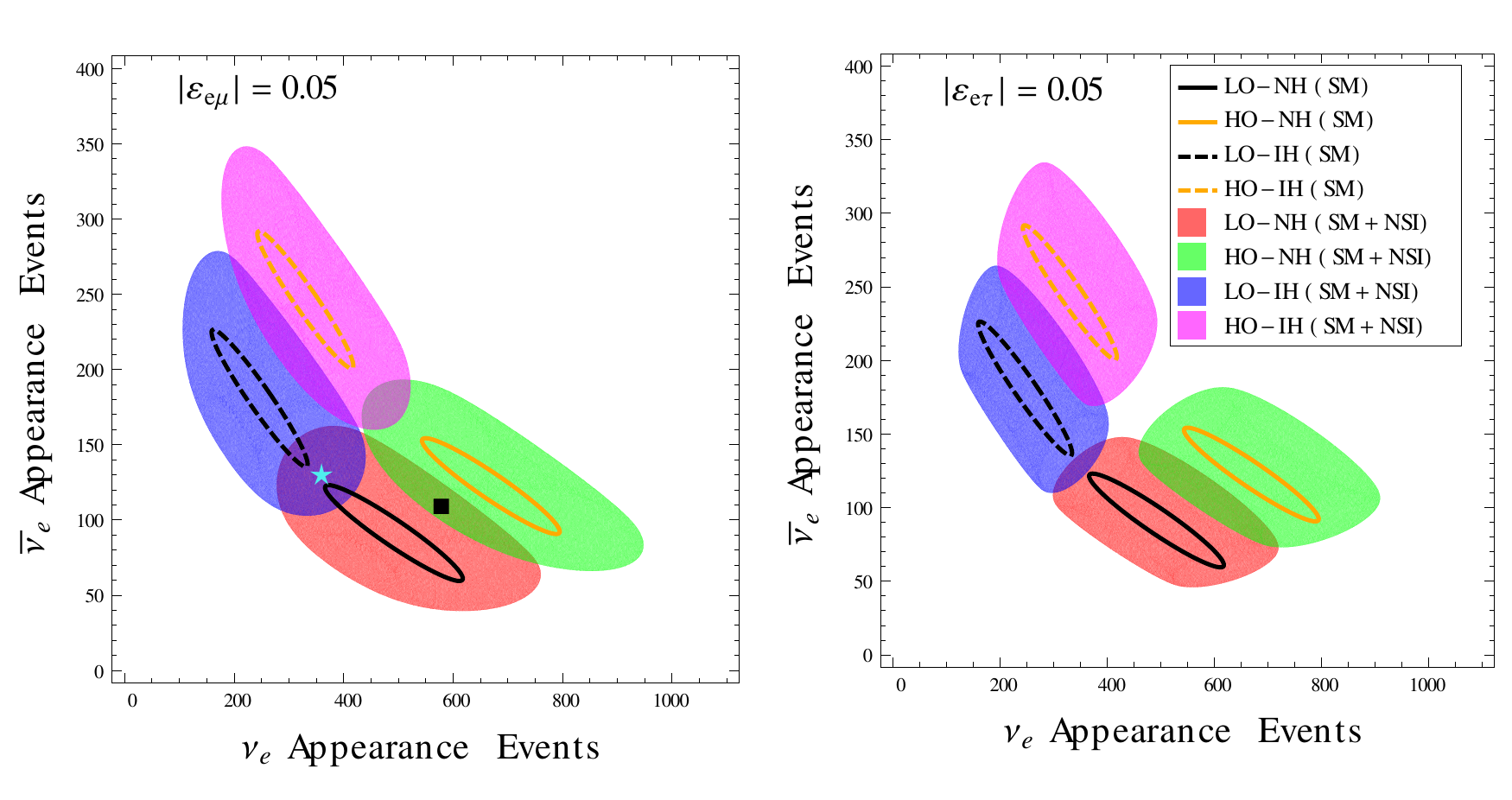}
\vspace{-0.2cm}
\caption{Bi-event plot for the DUNE setup. The ellipses represent
the SM case, while the colored blobs correspond to SM+NSI (see the legends). 
We take $\sin^2\theta_{23} = 0.42$\,(0.58) as  benchmark value for the LO (HO).
In the SM ellipses, the running parameter is $\delta$ varying in the range $[-\pi,\pi]$. 
In case of SM+NSI blobs, there are two running parameters: $\delta$ and $\phi_{e\mu}$ 
($\delta$ and $\phi_{e\tau}$) in the left (right) panel, both varying 
in their allowed ranges $[-\pi,\pi]$.}
\label{fig:bievents}
\end{figure}  

Figure~\ref{fig:bievents} provides a useful geometrical representation of the situation.   
In such a plot, the ellipses refer to the SM case, while the colored blobs represent the SM+NSI scheme.
In each panel we show the four cases corresponding to the different choice of the neutrino mass
hierarchy (MH), which can be normal (NH) or inverted (IH), and to the different choice of the octant (LO and HO).
We have taken $\sin^2 \theta_{23} = 0.42\,(0.58)$ as a benchmark value for the LO (HO) octant.
In the left (right) panel we have switched on the $e\mu$ ($e\tau$) coupling taking for its modulus 
$|\varepsilon_{e\mu}| = 0.05$ ($|\varepsilon_{e\tau}| = 0.05$) and varying the associated CP-phase
$\phi_{e\mu}$ ($\phi_{e\tau}$) in its allowed interval $[-\pi, \pi]$. 
The graphs neatly show that the $\theta_{23}$ octant separation existing in the SM case is lost in the presence of NSI's
since the two separate ellipses become overlapping blobs. We can understand how the blobs arise thinking them as 
a convolution of an infinite ensemble of ellipses (for more examples, see~\cite{Agarwalla:2016mrc,Agarwalla:2016xxa}), 
each corresponding to a different value of the new phase ($\phi_{e\mu}$ or $\phi_{e\tau}$). The orientation of the ellipses 
changes as a function of such new CP-phase covering a full area in the bi-event space. The shape of the colored blobs 
is slightly different between the two cases of  $\varepsilon_{e\mu}$ and $\varepsilon_{e\tau}$ as a result of the 
different functional dependency of the transition probability. One can notice that in both 
panels there is also an overlap among the two hierarchies, which is more pronounced in the $e\mu$ case (left panel)
if compared with the $e\tau$ case (right panel). This may indicate that the MH may be a source of 
degeneracy in the octant identification. 

\begin{figure}[t]
\vspace{-0.5cm}
\hspace{1.2cm}
\includegraphics[width=14.cm]{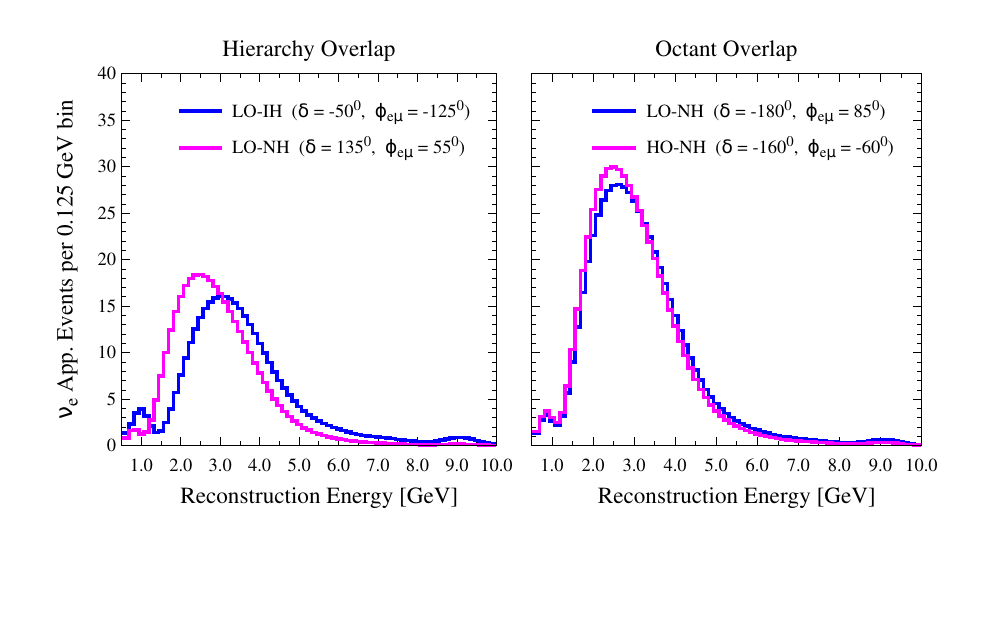}
\vspace{-1.5cm}
\caption{Electron neutrino spectra of DUNE for $|\varepsilon_{e\mu}|=0.05$ plotted for four representative cases.  
The left panel illustrates the comparison of two cases in which the total number of 
events is exactly the same for NH and IH (corresponding to the cyan star in the left panel of Fig.~\ref{fig:bievents}).
The right panel displays the comparison of two cases in which the total number of events is exactly the same for LO and HO
(corresponding to the black square in the left panel of Fig.~\ref{fig:bievents}). See the text for more details.}
\label{fig:spectrum}
\end{figure}  

This is not the case, however, because in the DUNE experiment the energy spectrum brings additional
information that breaks the MH degeneracy. In contrast, the energy spectrum is not able to offer much
help in lifting the octant degeneracy. This behavior is elucidated by Fig.~\ref{fig:spectrum}, 
which represents the reconstructed electron neutrino event spectra in DUNE 
for $|\varepsilon_{e\mu}|=0.05$ plotted for four representative cases.
The left panel illustrates the comparison of two cases in which the total number of $\nu_e$
events is exactly the same for NH and IH.  The right panel displays the comparison of two cases in 
which the total number of events is exactly the same for LO and HO. 
The two spectra on the left panel are calculated for the values
of the CP-phases $\delta$ and $\phi_{e\mu}$ indicated in the legend, which correspond to the same point 
in the bievent space located in the overlapping region of the two (red and blue) LO blobs (the cyan star in the 
left panel of Fig.~\ref{fig:bievents}). The two spectra on the right panel are calculated for the values
of the CP-phases $\delta$ and $\phi_{e\mu}$ indicated in the legend,  which correspond to the
same point in the bievent space located in the overlapping region of the two (red and green) NH blobs 
(the black square in the left panel of Fig.~\ref{fig:bievents}). Figure~\ref{fig:spectrum} clearly shows that, while
the spectra are rather different for the two hierarchies, especially at low energies, 
they are almost identical for the two octants. We find a similar behavior in the electron
antineutrino spectra (not shown) for the same choices of the CP-phases indicated in the legend
of Fig.~\ref{fig:spectrum}.  This implies that the MH is not a source of confusion in the 
octant identification%
\footnote{In~\cite{Liao:2016hsa,Coloma:2016gei}, it has been pointed out that if $\varepsilon_{ee}$ is
non-zero and $\mathcal{O}(1)$, then DUNE alone cannot determine the correct MH. 
In such a scenario, the octant ambiguity that we are dealing with can be further exacerbated.}.
Nonetheless, for generality, in the numerical analysis presented in the 
next section, we will treat the MH as an unknown parameter.

\section{Numerical results}

For DUNE, we consider a 35 kiloton fiducial liquid argon far detector in our work, and follow the
detector characteristics which are mentioned in Table 1 of Ref.~\cite{Agarwalla:2011hh}.
We assume a proton beam power of 708 kW in its initial phase with a proton energy of 
120 GeV which can deliver $6 \times 10^{20}$ protons on target in 230 days per calendar 
year. In our calculation, we have used the fluxes which were obtained assuming a decay pipe 
length of 200 m and 200 kA horn current~\cite{mbishai}. We take a total run time of ten years,
which is equivalent to a total exposure of 248 kiloton $\cdot$ MW $\cdot$ year,
equally shared between neutrino and antineutrino modes. In our work, we consider the 
reconstructed energy range of neutrino and antineutrino to be 0.5 GeV to 10 GeV. 
As far as the systematic uncertainties are concerned, we assume an uncorrelated 
5\% normalization error on signal, and 5\% normalization error on background for both 
the appearance and disappearance channels. The same set of systematics are taken for 
both the neutrino and antineutrino channels which are also uncorrelated. 
In our simulations, we use the GLoBES software~\cite{Huber:2004ka,Huber:2007ji}.
We incorporate the effect of the NSI parameters both in the $\nu_\mu \to \nu_e$ 
appearance channel, and in the $\nu_\mu \to \nu_\mu$ disappearance channel. 
The same is also applicable for the antineutrino run. 
The benchmark (central) values of the three-flavor oscillation parameters that we consider in this work are: 
$\sin^2\theta_{12}$ = 0.304, $\sin^2 2\theta_{13}$ = 0.085, $\sin^2\theta_{23}$ = 0.42 (0.58) for LO (HO),
$\Delta m_{21}^2$ = $7.50 \times 10^{-5}$ eV$^2$, $\Delta m_{31}^2$ (NH) = $2.475 \times 10^{-3}$ eV$^2$,
$\Delta m_{31}^2$ (IH) = - $2.4 \times 10^{-3}$ eV$^2$, and the CP phase $\delta$ in the range [-$\pi$, $\pi$].
These choices of the oscillation parameters are in close agreement with the recent best-fit values from
Ref.~\cite{Capozzi:2016rtj,Gonzalez-Garcia:2015qrr,Forero:2014bxa}. For the cases, where the results are
shown as a function of true value of $\sin^2\theta_{23}$, we consider the 3$\sigma$ allowed range of 
0.38 to 0.63. For the DUNE baseline of 1300 km, we take the the line-averaged constant Earth matter 
density of $\rho=2.87\,\mathrm{g/cm^3}$ estimated using the Preliminary Reference Earth Model 
(PREM)~\cite{PREM:1981}. To obtain the numerical results, we carry out a full spectral 
analysis using the binned events spectra for DUNE. 
In order to determine the sensitivity  of DUNE for excluding the false octant, we define the Poissonian $\Delta \chi^2$ as
\begin{eqnarray}
\Delta\chi^2 = \chi^2_{\mathrm{false\,\, octant}}  - \chi^2_{\mathrm{true\,\, octant}}\,,
\label{eq:chi2}
\end{eqnarray}
where $\chi^2_{\mathrm{true\,\, octant}}$ ($\chi^2_{\mathrm{false\,\, octant}}$) is generated for 
the true (test) values of ($\theta_{23},\delta,\phi$). To obtain the curves displayed in Fig.~3, for any given 
choice of the true parameters, we minimize the $\Delta\chi^2$ in Eq.~(\ref{eq:chi2}) with respect to the test
parameters varying $\theta_{23}^{\mathrm{test}}$ in the false octant and ($\delta^{\mathrm{test}}, \phi^{\mathrm{test}}$)
 in the range $[-\pi,\pi]$.
In addition, in Fig.~4 we marginalize also over $\phi^{\mathrm{true}}$ in the range $[-\pi,\pi]$.
Finally, in Fig.~5 with also marginalize over $\delta^{\mathrm{true}}$. We follow the method of pulls as described in 
Refs.~\cite{Huber:2002mx,Fogli:2002pt} to marginalize  $\Delta\chi^{2}$ 
over the uncorrelated systematic uncertainties. To give our results at $1,2,3\sigma$ 
confidence levels for 1 d.o.f., we use the relation $\textrm{N}\sigma \equiv \sqrt{\Delta\chi^2}$,
which is valid in the frequentist method of hypothesis testing~\cite{Blennow:2013oma}.

\begin{figure*}[t]
\vspace*{-1.1cm}
\hspace*{1.3cm}
\includegraphics[width=13.5 cm]{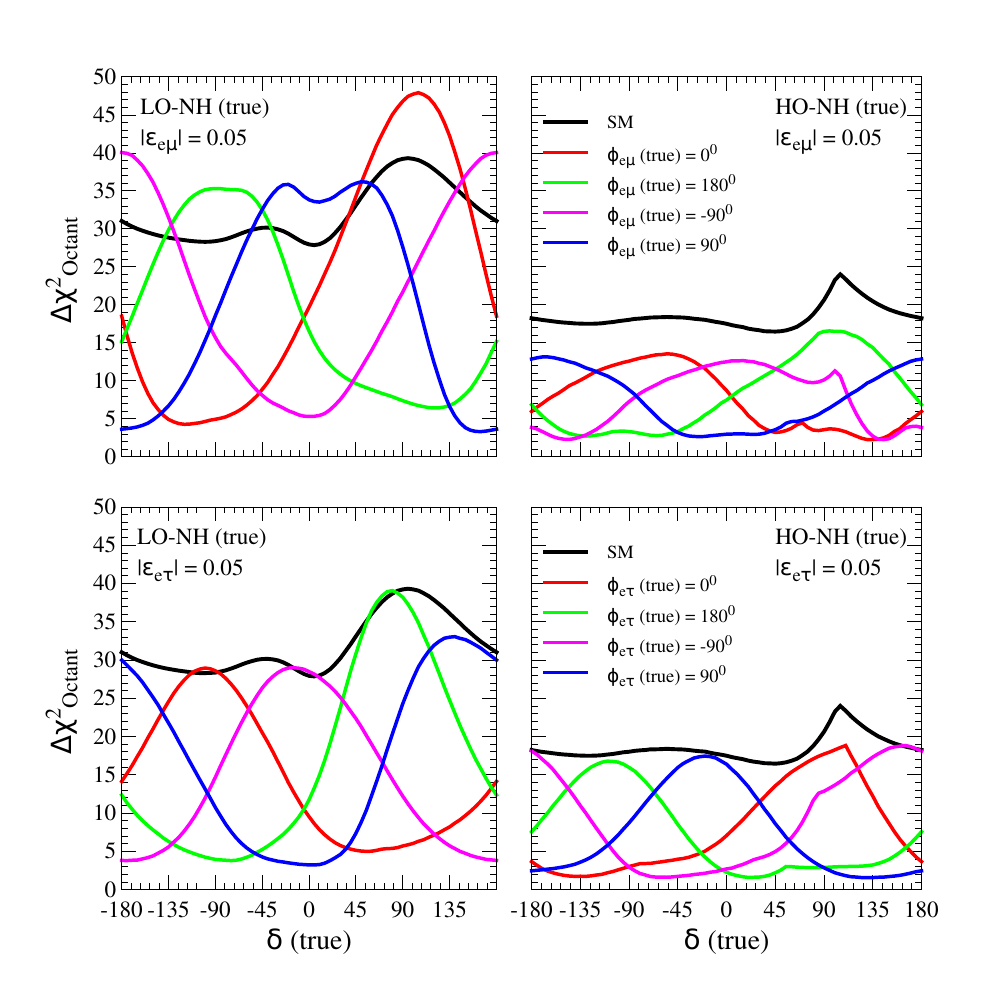}
\vspace*{-0.5cm}
\caption{Discovery potential of the true octant as a function of true $\delta$ 
assuming LO-NH (left panels) and HO-NH (right panels) as the true choice. 
We take $\sin^2\theta_{23} = 0.42$\,(0.58) as  benchmark value for the LO (HO). 
In each panel, we present the results for the SM case (black line), and for the 
SM+NSI scheme (colored lines) considering four different values of true $\phi_{e\mu}$
(upper panels) and $\phi_{e\tau}$ (lower panels). In the SM case, we marginalize away 
($\theta_{23}, \delta $) (test). In the SM+NSI scheme, we fix $|\varepsilon_{e\mu}| = 0.05$ 
in the two upper panels and  $|\varepsilon_{e\tau}| = 0.05$ in the two lower panels.
In the two upper panels, we marginalize over ($\theta_{23}, \delta, \phi_{e\mu}$) (test), while in the
two lower ones, we marginalize over ($\theta_{23}, \delta, \phi_{e\tau}$) (test).}
 \label{fig:2pan_octant_sens}
\end{figure*}  
 
Figure~\ref{fig:2pan_octant_sens} displays the sensitivity for excluding the wrong octant as a function of true $\delta$. The two upper panels refer to $\varepsilon_{e\mu}$ while the two lower panels refer to $\varepsilon_{e\tau}$. In each case we fix the modulus of the coupling ($|\varepsilon_{e\mu}|$ or $|\varepsilon_{e\tau}|$) equal to 0.05. The two left (right) panels refer to the true choice LO-NH (HO-NH). In all panels, for the sake of comparison, we show the results for the 3-flavor SM case (represented by the black curve). Concerning the SM+NSI scheme, we draw the curves corresponding to four representative values of the (true) dynamical CP-phase ($\phi_{e\mu}$ or $\phi_{e\tau}$). In the SM case we have marginalized over ($\theta_{23}, \delta$) (test). In the SM+NSI scheme, we have also marginalized over the test value of the new dynamical CP-phase ($\phi_{e\mu}$ or $\phi_{e\tau}$). In all cases we have marginalized over the mass hierarchy. However, we have checked that the minimum of $\Delta \chi^2$ is never reached in the wrong hierarchy. This confirms that the neutrino mass hierarchy is not an issue in the determination of the $\theta_{23}$ octant in DUNE, as expected on the basis of the discussion about the energy spectral information made in the previous section.

\begin{figure*}[t]
\vspace{-0.4cm}
\hspace{0.05cm}
\includegraphics[width=1\textwidth]{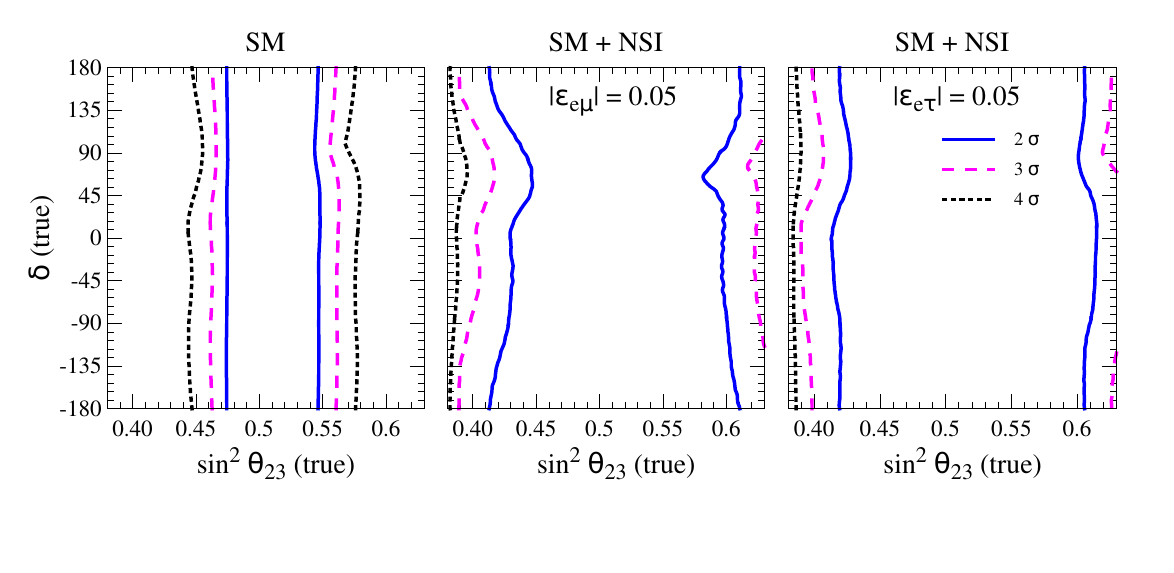}
\vspace*{-1.7cm}
\caption{Discovery potential of the true octant in [$\sin^2 \theta_{23}, \delta$] (true) plane
assuming NH as true choice. The left panel corresponds to the SM case. The middle (right)
panel represents the SM+NSI case where we have switched on $\varepsilon_{e\mu}$ ($\varepsilon_{e\tau}$).
In the SM case, we marginalize away ($\theta_{23}, \delta$) (test). In the SM+NSI cases, 
in addition, we marginalize over the true and test value of the additional CP-phase ($\phi_{e\mu}$ 
in the middle panel, $\phi_{e\tau}$ in the right panel). The solid blue, 
dashed magenta, and dotted black curves correspond, respectively, to the 
2$\sigma$, 3$\sigma$, and 4$\sigma$ confidence levels (1 d.o.f.).}
\label{fig:3pan_octant_true}
\end{figure*}

\begin{figure}[t]
\vspace{-1.1cm}
\hspace{0.3cm}
\includegraphics[width=16cm]{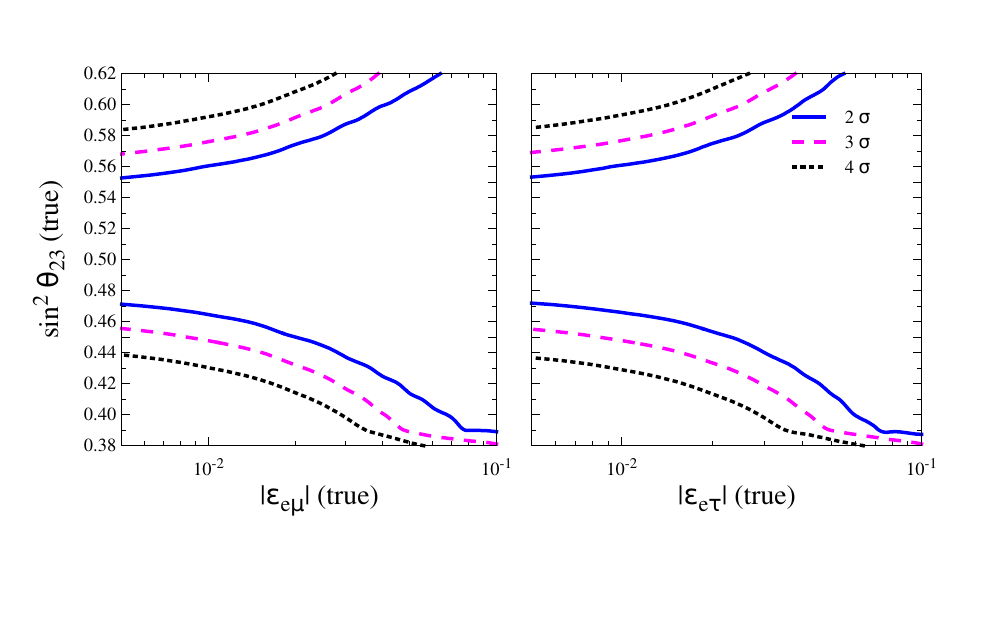}
  \label{bievents-convolution}
  \vspace{-1.7cm}
  \caption{Degradation of the $\theta_{23}$ octant sensitivity as a function of the NSI strength $|\varepsilon|$,
assuming NH as true choice. The left (right) panel corresponds to 
$\varepsilon  \equiv \varepsilon_{e\mu}$ ($\varepsilon \equiv \varepsilon_{e\tau}$).
In both cases, $\theta_{23}$ (test) and  $\delta$ (both true and test) have been marginalized away. 
In addition, in the left (right) panel, the true and test values of the CP-phase $\phi_{e\mu}$  ($\phi_{e\tau}$)
have been marginalized away. The solid blue, dashed magenta, and dotted black curves correspond, respectively, to the 
2$\sigma$, 3$\sigma$, and 4$\sigma$ confidence levels (1 d.o.f.).}
\label{fig:2pan_octant_general}
 \end{figure}

In the analysis shown in Fig~\ref{fig:2pan_octant_sens}, we have fixed 
$\sin^2\theta_{23} = 0.42$\,(0.58) as a  benchmark value for the LO (HO), corresponding to 
a deviation $\eta =\pm 0.08$. In general, one may want to know how things change for different choices
of the true value of $\theta_{23}$ since it is unknown. Figure~\ref{fig:3pan_octant_true} gives a quantitative
answer to this question. It displays the discovery potential of the true octant in the [$\sin^2 \theta_{23}, \delta$] (true) plane,
assuming NH as true choice. The left panel corresponds to the SM case. The middle (right)
panel represents the SM+NSI case, where we have ``switched on'' $\varepsilon_{e\mu}$ ($\varepsilon_{e\tau}$)
with modulus 0.05. In the SM case we have marginalized away ($\theta_{23}, \delta$) (test). In the SM+NSI cases, 
in addition, we have marginalized over the true and test value of the new dynamical CP-phase ($\phi_{e\mu}$ 
in the middle panel, $\phi_{e\tau}$ in the right panel). The solid blue,  dashed magenta, and dotted black curves correspond,
respectively, to the 2$\sigma$, 3$\sigma$, and 4$\sigma$ confidence levels (1 d.o.f.). From the comparison
of the middle and right panels with the left one, we can see that the presence of NSI with strength $|\varepsilon| = 0.05$
compromises the octant sensitivity for all the phenomenologically interesting region allowed for $s_{23}^2$ 
by current data.  This is of particular interest because such low strengths of the NSI's are well below
the current upper bounds both for $\varepsilon_{e\mu}$ and $\varepsilon_{e\tau}$. Finally, it is interesting
to ask how the deterioration of the $\theta_{23}$ octant discovery potential varies with the NSI strength.
For this purpose one needs to treat the NSI strength as a free parameter, allowing the associated CP-phase
to vary in the interval $[-\pi,\pi]$. The results of this general analysis are represented in Fig~\ref{fig:2pan_octant_general},
which shows the discovery potential of the $\theta_{23}$ octant in the plane  [$|\varepsilon|, \sin^2 \theta_{23}$] (true),
assuming NH as true choice.  The left (right) panel corresponds to 
$\varepsilon  \equiv \varepsilon_{e\mu}$ ($\varepsilon \equiv \varepsilon_{e\tau}$).
In both cases, the standard parameters ($\theta_{23}, \delta$) (test) and $\delta$ (true) have been marginalized away. 
In addition, in the left (right) panel the true and test values of the CP-phase $\phi_{e\mu}$  ($\phi_{e\tau}$)
have been marginalized away. We observe that for NSI strengths below the 1\% level, the sensitivity substantially
coincides with that achieved in the SM case. In this case the NSI's are harmless. For larger values, 
the sensitivity gradually deteriorates, until it basically
goes below the $2\sigma$ level for all the interesting values of $\sin^2 \theta_{23}$ if $|\varepsilon| \gtrsim 0.07$.

Before concluding this section, a remark is in order concerning the off-diagonal coupling $\varepsilon_{\mu\tau}$. First,
one should note that this coupling is the most strongly constrained due to the high sensitivity of
atmospheric neutrino data to the $\nu_\mu \to \nu_\tau$ transitions. The most recent Super-Kamiokande 
analysis provides the upper bound $|\varepsilon_{\mu\tau}| \lesssim \mathrm {0.033}$ at the 90\% C.L.~\cite{Mitsuka:2011ty} (see also~\cite{GonzalezGarcia:2011my}), whose results are corroborated by MINOS data~\cite{Adamson:2013ovz}. Second, in the context of long-baseline experiments, $\varepsilon_{\mu\tau}$ essentially affects only the $\nu_\mu \to \nu_\mu$ disappearance probability, while its effects on the $\nu_\mu \to \nu_e$ appearance probability are negligible. These two circumstances make the $\varepsilon_{\mu\tau}$ coupling less important for what concerns the discrimination of the octant of $\theta_{23}$. This fact is corroborated by our numerical simulations. We have explicitly verified that even for $|\varepsilon_{\mu\tau}| = 0.05$, which is well above the present upper bound, the DUNE sensitivity to the octant of $\theta_{23}$ never goes below 4.4$\sigma$ (3.1$\sigma$) for the benchmark value $s^2_{23} = 0.42\,(0.58)$ for LO\,(HO). Also, we find only mild changes in the sensitivity when the associated CP-phase $\phi_{\mu\tau}$ is allowed to vary in the interval $[-\pi,\pi]$.

\section{Conclusions}

We have investigated the impact of non-standard flavor-changing interactions (NSI) on the reconstruction of the octant of the atmospheric mixing angle $\theta_{23}$ in the next generation LBL experiments, taking the Deep Underground Neutrino Experiment (DUNE) as a case study. In the presence of such new interactions the $\nu_\mu \to \nu_e$  transition probability acquires an additional interference term, which depends on one new dynamical CP-phase $\phi$. This term sums up with the well-known interference term related to the standard CP-phase $\delta$. For values of the NSI coupling as small as $few\,\%$  (relative to the Fermi constant $G_{\mathrm F}$) the combination of the two interference terms can mimic a swap of the $\theta_{23}$ octant. As a consequence, for unfavorable values of the two CP-phases $\delta$ and $\phi$, the discovery potential of the octant of $\theta_{23}$ gets completely lost.  We point out that the degeneracy between the octant of $\theta_{23}$ and NSI's discussed in this paper has now become more important in light of the new results from the NO$\nu$A Collaboration presented a few days ago at the Neutrino 2016 conference, which suggest that maximal $\theta_{23}$ is disfavored
at the $2.5 \sigma$ confidence level~\cite{NOvA:2016}. 

We close the paper with a general remark. In a previous work~\cite{Agarwalla:2016xlg},
we found that a similar loss of sensitivity to the $\theta_{23}$ octant can occur due to the presence of a light eV-scale sterile neutrino. Also in that case a new interference term appears in the $\nu_\mu \to \nu_e$ transition probability, which depends
on one additional CP-phase.  Therefore, albeit in the two cases the origin of the new CP-phase is completely different, having kinematical nature in the sterile neutrino case and dynamical nature in the NSI case, their phenomenological manifestation at the far detector of LBL experiments is very similar. On the basis of this observation, we can predict an analogous behavior also for other mechanisms which involve a new interference term in the transition probability, like for example the violation of unitarity of the PMNS matrix recently investigated in~\cite{Miranda:2016wdr}. Therefore, we can conclude that in general, 
whenever a new interference term due to any new physics crops up in the LBL $\nu_\mu \to \nu_e$ appearance probability,
the reconstruction of the $\theta_{23}$ octant may be in danger.

\section*{Acknowledgments}

S.K.A. is supported by the DST/INSPIRE Research Grant [IFA-PH-12],
Department of Science \& Technology, India. A.P. is supported by the 
Grant ``Future In Research'' {\it Beyond three neutrino families},
contract no. YVI3ST4, of Regione Puglia, Italy. 

\section*{References}

\bibliographystyle{apsrev-title}

\bibliography{NSI-References}  

\end{document}